\documentclass[12pt]{iopart}
\newcommand{\EQ}{\begin{equation}}
\newcommand{\EN}{\end{equation}}
\newcommand{\EQA}{\begin{eqnarray}}
\newcommand{\ENA}{\end{eqnarray}}
\newcommand{\eq}[1]{(\ref{#1})}
\newcommand{\EEq}[1]{Equation~(\ref{#1})}
\newcommand{\Eq}[1]{equation~(\ref{#1})}
\newcommand{\Eqs}[2]{equations~(\ref{#1}) and~(\ref{#2})}

\newcommand{\Sec}[1]{\S\ref{#1}}

\newcommand{\Fig}[1]{Fig.~\ref{#1}}

\newcommand{\Tab}[1]{Table~\ref{#1}}
\newcommand{\Figs}[2]{Figs~\ref{#1} and \ref{#2}}

\newcommand{\bra}[1]{\langle #1\rangle}

{}
\newcommand{\meanFF}{\overline{\mbox{\boldmath ${\cal F}$}}{}}{}

{}
{}
{}
\newcommand{\meanEMF}{\overline{\mbox{\boldmath ${\cal E}$}}{}}{}
{}
{}
{}
{}
\newcommand{\meanBB}{\overline{\mbox{\boldmath $B$}}{}}{}
\newcommand{\meanJJ}{\overline{\mbox{\boldmath $J$}}{}}{}
\newcommand{\meanUU}{\overline{\mbox{\boldmath $U$}}{}}{}
{}

\newcommand{\meanB}{\overline{B}}

\newcommand{\meanU}{\overline{U}}

\newcommand{\hatBB}{\hat{\mbox{\boldmath $B$}}{}}{}
\newcommand{\hatJJ}{\hat{\mbox{\boldmath $J$}}{}}{}
\newcommand{\hatEMF}{\hat{\mbox{\boldmath ${\cal E}$}}{}}{}
\newcommand{\eee}{\hat{\mbox{\boldmath $e$}} {}}

\newcommand{\zzz}{\hat{\mbox{\boldmath $z$}} {}}

\newcommand{\kk}{\mbox{\boldmath $k$} {}}

\newcommand{\uu}{\mbox{\boldmath $u$} {}}
\newcommand{\UU}{\mbox{\boldmath $U$} {}}
\newcommand{\xx}{\mbox{\boldmath $x$} {}}

\newcommand{\BB}{\mbox{\boldmath $B$} {}}
\newcommand{\EE}{\mbox{\boldmath $E$} {}}

\newcommand{\JJ}{\mbox{\boldmath $J$} {}}

\newcommand{\AAA}{\mbox{\boldmath $A$} {}}

\newcommand{\ff}{\mbox{\boldmath $f$} {}}

\newcommand{\EEE}{\mbox{\boldmath ${\cal E}$} {}}
\newcommand{\FFF}{\mbox{\boldmath ${\cal F}$} {}}

\newcommand{\WW}{\mbox{\boldmath $W$} {}}
\newcommand{\QQ}{\mbox{\boldmath $Q$} {}}

\newcommand{\nab}{\mbox{\boldmath $\nabla$} {}}

\newcommand{\RRRR}{\mbox{\boldmath ${\sf R}$} {}}
\newcommand{\SSSS}{\mbox{\boldmath ${\sf S}$} {}}

\newcommand{\ii}{{\rm i}}
\newcommand{\erf}{{\rm erf}}

\newcommand{\DD}{{\rm D} {}}

\newcommand{\dd}{{\rm d} {}}
\newcommand{\const}{{\rm const}  {}}

\def\half{{\textstyle{1\over2}}}

\def\onethird{{\textstyle{1\over3}}}

\newcommand{\yapj}[3]{ #1, {ApJ,} {#2}, #3}

\newcommand{\yapjl}[3]{ #1, {ApJ,} {#2}, #3}

\newcommand{\yan}[3]{ #1, {Astron.\ Nachr.,} {#2}, #3}

\newcommand{\yana}[3]{ #1, {A\&A,} {#2}, #3}

\newcommand{\ygafd}[3]{ #1, {Geophys.\ Astrophys.\ Fluid Dyn.,} {#2}, #3}
\newcommand{\ygrl}[3]{ #1, {Geophys.\ Res.\ Lett.,} {#2}, #3}

\newcommand{\yjfm}[3]{ #1, {J.\ Fluid Mech.,} {#2}, #3}

\newcommand{\ypf}[3]{ #1, {Phys.\ Fluids,} {#2}, #3}
\newcommand{\ypp}[3]{ #1, {Phys.\ Plasmas,} {#2}, #3}

\newcommand{\yprl}[3]{ #1, {Phys.\ Rev.\ Lett.,} {#2}, #3}

\newcommand{\ymn}[3]{ #1, {MNRAS,} {#2}, #3}
\newcommand{\ynat}[3]{ #1, {Nature,} {#2}, #3}

\newcommand{\ysph}[3]{ #1, {Solar Phys.,} {#2}, #3}

\newcommand{\ypre}[3]{ #1, {Phys.\ Rev.\ E,} {#2}, #3}

\newcommand{\yjour}[4]{ #1, {#2}, {#3}, #4}

\usepackage{graphicx,url}
\usepackage[normalem]{ulem}
\begin{document}

\title[Magnetic helicity effects in astrophysical and laboratory dynamos]
{Magnetic helicity effects in astrophysical and laboratory dynamos}

\author{A Brandenburg$^1$ and P J K\"apyl\"a$^{1,2}$}
\address{1. NORDITA, AlbaNova University Center, SE-10691 Stockholm, Sweden}
\address{2. Observatory, P.O. Box 14, FI-00014 University of Helsinki, Finland}

\ead{brandenb@nordita.dk}
\begin{abstract}
Magnetic helicity effects are discussed in laboratory and astrophysical
settings.
First, dynamo action in Taylor-Green flows is discussed for different
boundary conditions.
However, because of the lack of scale separation with respect to the
container, no large scale field is being produced and there is no
resistively slow saturation phase as otherwise expected.
Second, the build-up of a large scale field is demonstrated in
a simulation where a localized magnetic eddy produces field on
a larger scale if the eddy possesses a swirl.
Such a set-up might be realizable experimentally through coils.
Finally, new emerging issues regarding the connection between
magnetic helicity and the solar dynamo are discussed.
It is demonstrated that dynamos with a non-local (Babcock-Leighton
type) alpha effect can also be catastrophically quenched, unless
there are magnetic helicity fluxes.

\end{abstract}


\section{Introduction}

Many astrophysical dynamos are driven by helical flows via an $\alpha$ effect.
Helical flows are also employed in all laboratory realizations of liquid
metal dynamos.
One of the remarkable properties of helical dynamos is that they produce
large scale fields that are helical.
Since net helicity is conserved, this can only happen if here is a production
of an equal amount of small scale fields with opposite helicity.
This has been demonstrated through various numerical studies
(Brandenburg 2001, Mininni et al.\ 2005a).

An obvious question is whether these properties can also be seen in
experimental realization of fluid dynamos.
The preliminary answer to this is no, because there is no scale separation.
We will return to this in more detail and present new calculations of
Taylor-Green flows that model the flow in the French VKS2 experiment
(Monchaux et al.\ 2007).
To address the issue of the lack of scale separation we also study a model
designed to show the development of large scale magnetic fields
that are driven from a small localized source.
Finally, we discuss some new issues regarding the solar dynamo.

We assume some level of familiarity with the concept of magnetic helicity
conservation and the resistively limited saturation phenomenon found in the
nonlinear evolution of large scale dynamos.
In this connection we highlight the usage of the word ``catastrophic'', which
is meant to indicate that the result depends on the value of the
magnetic Reynolds number.
This applies in particular to the $\alpha$ effect which is now known to be
catastrophically quenched when there is no flux of magnetic helicity.
A recent review on the development of the last five years can be found in
Brandenburg \& Subramanian (2005a).

Throughout this paper we present original results that have not yet been
presented earlier.
Many of the cases considered are motivated by the recent developments
in laboratory dynamos and solar dynamo modeling.
Many of the simulations have been carried out at relatively low resolution.
The present results are therefore tentative and need to be followed up using
higher resolution simulations.
However, the results presented here do reflect the current state of
affairs in this field, which is indeed one of the objectives of this paper.

\section{Governing equations}

We consider here the isothermal and weakly compressible case, which
means that the pressure gradient term can be written as a gradient of
the pseudo enthalpy,\footnote{For a polytropic gas the enthalpy can be
written as $h=(\gamma-1)^{-1}c_{\rm s0}^2(\rho/\rho_0)^{\gamma-1}$, which
reduces to $c_{\rm s0}^2\ln(\rho/\rho_0)+\const$ in the limit of $\gamma\to1$.
In the following we include the above constant in our definition of
the enthalpy and refer to it therefore as the pseudo enthalpy.}
$h=c_{\rm s}^2\ln\rho$, where $c_{\rm s}=\const$ is
the isothermal speed of sound and $\rho$ is the density.
The induction equation can then be written in an analogous form
in terms of the magnetic vector potential by using the pseudo Lorenz
gauge with a freely specified speed $c_\phi$, which is for practical
reasons less than the speed of light, and may well be equal to the
speed of sound $c_{\rm s}$.
The set of equations is thus
\EQA
{\partial\AAA\over\partial t}=
-\nab\phi+\EEE_{\rm ext}-\eta\JJ+\uu\times\BB,
&\quad&{\partial\phi\over\partial t}=-c_\phi^2\nab\cdot\AAA,
\label{eqn1}
\\
{\DD\UU\over\DD t}=
-\nab h+\FFF_{\rm ext}-\nu\QQ+\JJ\times\BB/\rho,
&\quad&{\DD h\over\DD t}=-c_{\rm s}^2\nab\cdot\UU,
\label{eqn2}
\ENA
where $\UU$ is the velocity, $\BB=\nab\times\AAA$ is the magnetic field,
expressed in terms of the magnetic vector potential $\AAA$,
$\JJ=\nab\times\nab\times\AAA$ is the current density,
$\QQ=\nab\times\nab\times\UU$ is the double curl of the velocity,
$\DD/\DD t=\partial/\partial t+\UU\cdot\nab$ is the advective derivative,
and $\EEE_{\rm ext}$ and $\FFF_{\rm ext}$ are external forcing
functions, to be specified later.
The current density is measured in units where the vacuum permeability
is unity.
The main analogy we want to stress here is that between the electric
potential $\phi$ in the uncurled induction equation and the pseudo
enthalpy $h$ in the momentum equation.
We should point out that, in order to not disturb the analogy between
the two equations, we have ignored a correction term in
\Eq{eqn1} where $\QQ$ has to be replaced by
\EQ
\QQ\to\QQ+\nab\nab\cdot\UU+\SSSS\cdot\nab\ln(\rho\nu),
\label{Qdef}
\EN
where ${\sf S}_{ij}=\half(U_{i,j}+U_{j,i})-\onethird\delta_{ij}\nab\cdot\UU$
is the traceless rate of strain tensor.
For incompressible flows these correction terms vanish, and they are small
for weakly compressible (small Mach number) flows considered here.
Nevertheless, in the numerical computations these extra terms in \Eq{Qdef}
are always included.

If we think of the velocity field being primarily just the vector
potential for the vorticity (see \ref{analogy}), then we can regard
the continuity equation, $\DD h/\DD t=-c_{\rm s}^2\nab\cdot\UU$,
as the gauge condition for the velocity in the Lorenz-like gauge.
In this sense we can interpret the Lorenz gauge condition for $\AAA$
as a continuity equation for $\phi$.
Note, however, that $\AAA$ is not a physically measurable quantity.

\section{Lorenz versus Weyl gauge}

Magnetic and electric fields are invariant under the
gauge transformation
\EQ
\AAA' = \AAA+\nab\Lambda,
\label{gauge_trans}
\EN
\EQ
\phi' = \phi-{\partial\Lambda\over\partial t}.
\label{gauge_trans_phi}
\EN
For numerical purposes it is often convenient to choose the gauge
$\Lambda=\int\phi\,\dd t$, which implies that $\phi'=0$.
Thus, instead of \Eq{eqn1} one solves the equation
$\partial\AAA'/\partial t=-\EE$.
The latter is usually referred to as the Weyl gauge.

\begin{figure}[t!]\begin{center}
\includegraphics[width=.9\columnwidth]{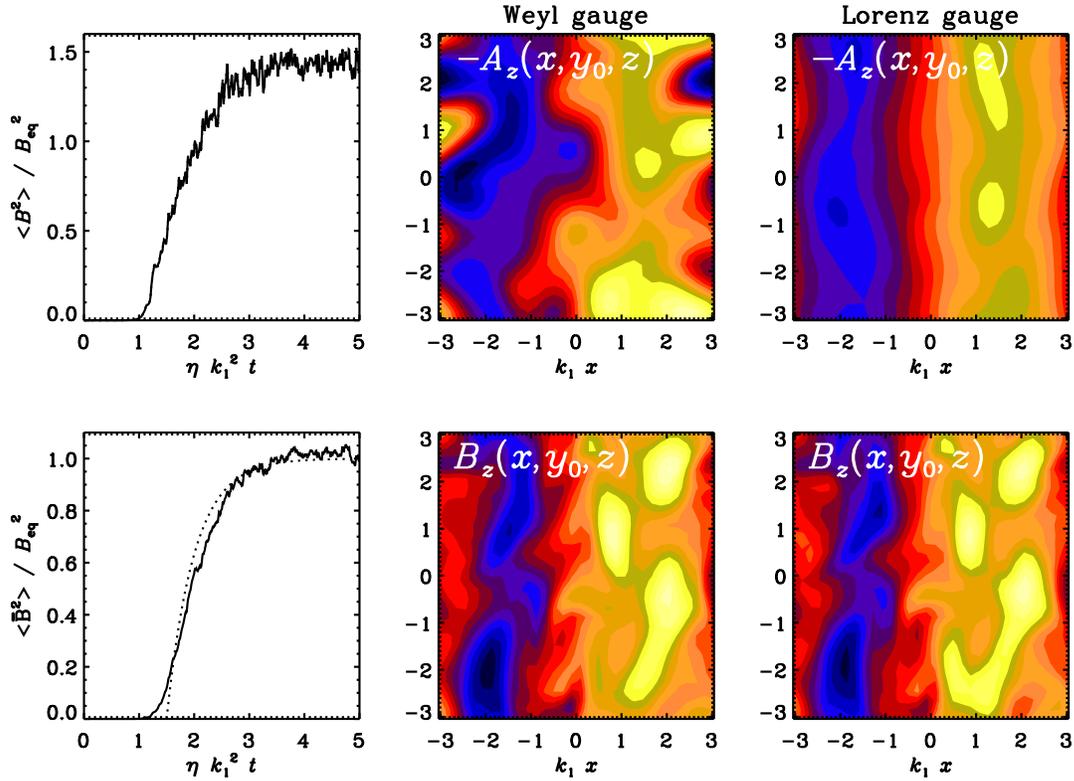}
\end{center}\caption[]{
Comparison of runs in Weyl and Lorenz gauges.
In this simulation a large scale magnetic field develops owing to the
helicity of the velocity forcing function.
While the magnetic field is the same in both gauges (second and third column
in the second row) and vector potential is different (second and third column
in the first row).
These slices are taken after 3 magnetic diffusion times.
For comparison, the evolution of magnetic energy in the total field
(upper left panel) and in the mean field (lower left panel) are shown,
where time is given in diffusion times.
}\label{pWeylLorenz}\end{figure}

In this section we compare the two gauges.
Obviously the magnetic field is the same in both cases,
but the magnetic vector potential is not.
In \Fig{pWeylLorenz} we compare the results obtained with the two
techniques in the case of a dynamo driven by fully helical turbulence.
The detailed forcing function used here is given in \Sec{ForcingFunction}.
We choose $32^3$ mesh points, a forcing wavenumber of $k_{\rm f}=3k_1$,
where $k_1$ is the smallest wavenumber in the box, and a forcing
amplitude of $0.07$, which results in an rms velocity of about $0.2$.
Viscosity and magnetic diffusivity are chosen to be $5\times10^{-3}$, so the
magnetic Reynolds number is $R_{\rm m}=u_{\rm rms}/(\eta k_{\rm f})=13$.
For details about these and similar runs at larger magnetic Reynolds
number, see Brandenburg (2001), which is also where the resistively
limited saturation phase with
\EQ
\bra{\meanBB^2}\propto 1-e^{-2\eta k_1^2(t-t_{\rm sat})}
\label{SlowSat}
\EN
was proposed.
Here, $t_{\rm sat}$ is the time where the small scale magnetic field
saturates.
The dotted line in the lower left panel shows this behavior.

For the present studies we used just $32^3$ meshpoints, which can
easily run on one processor using the \textsc{Pencil Code}\footnote{
\url{http://www.nordita.dk/software/pencil-code}.}, which is a
non-conservative, high-order, finite-difference code (sixth order in
space and third order in time) for solving the compressible hydrodynamic
equations.
The particular run presented in this section is actually one of the
sample runs ({\sf helical-MHDturb}) that come with the code.
Here and in all other cases presented in this paper we use
$c_{\rm s}=c_\phi=1$.

Obviously, the resistive saturation phase becomes more prominent at
larger Reynolds numbers.
However, the point of this simulation was to illustrate the differences
between Lorenz and Weyl gauges.
Qualitatively, it appears that the magnetic vector potential in the
Lorenz gauge is smoother than that in the Weyl gauge.
In general, the divergence of the magnetic vector potential is small,
so for all practical purposes the Lorenz gauge is close to the
Coulomb gauge.
One may hope that in the presence of open boundary conditions,
where a gauge-invariant magnetic helicity is harder to define,
the Lorenz-gauged magnetic vector potential may provide some meaningful
guidance regarding the escape of magnetic helicity density and the
associated magnetic helicity fluxes (cf.\ Subramanian \& Brandenburg 2006).

\section{Taylor-Green flow dynamos}

The Taylor-Green (TG) flow is often studied in connection with the
von K\'arm\'an Sodium (VKS) dynamo experiment in Cadarache in France
(Monchaux et al.\ 2007).
The TG flow is given by $\uu=\FFF_{\rm ext}/(\nu k_{\rm f}^2)$, where
$k_{\rm f}=\sqrt{3}k_0$ and
\EQ
\FFF_{\rm ext}= 2 f_0\pmatrix{
+\sin k_0 x \cos k_0 y \cos k_0 z\cr
-\cos k_0 x \sin k_0 y \cos k_0 z\cr
0},
\quad -{\pi\over k_0}< x,y,z <{\pi\over k_0}.
\EN
The forcing function is normalized such that $\bra{\FFF_{\rm ext}^2}^{1/2}=f_0$.
There is a vast literature on this flow.
Due to a large number of symmetries allowing computational simplifications,
large Reynolds numbers can be achieved.
This flow has therefore been used to study singularities and turbulence
(Nakano 1985, Brachet 1991). 
In more recent years dynamo action for this flow has been studied
(Nore et al.\ 1997), especially at low magnetic Prandtl numbers
(Ponty et al.\ 2004, 2005, Mininni et al.\ 2005b).

\begin{figure}[t!]\begin{center}
\includegraphics[width=.8\columnwidth]{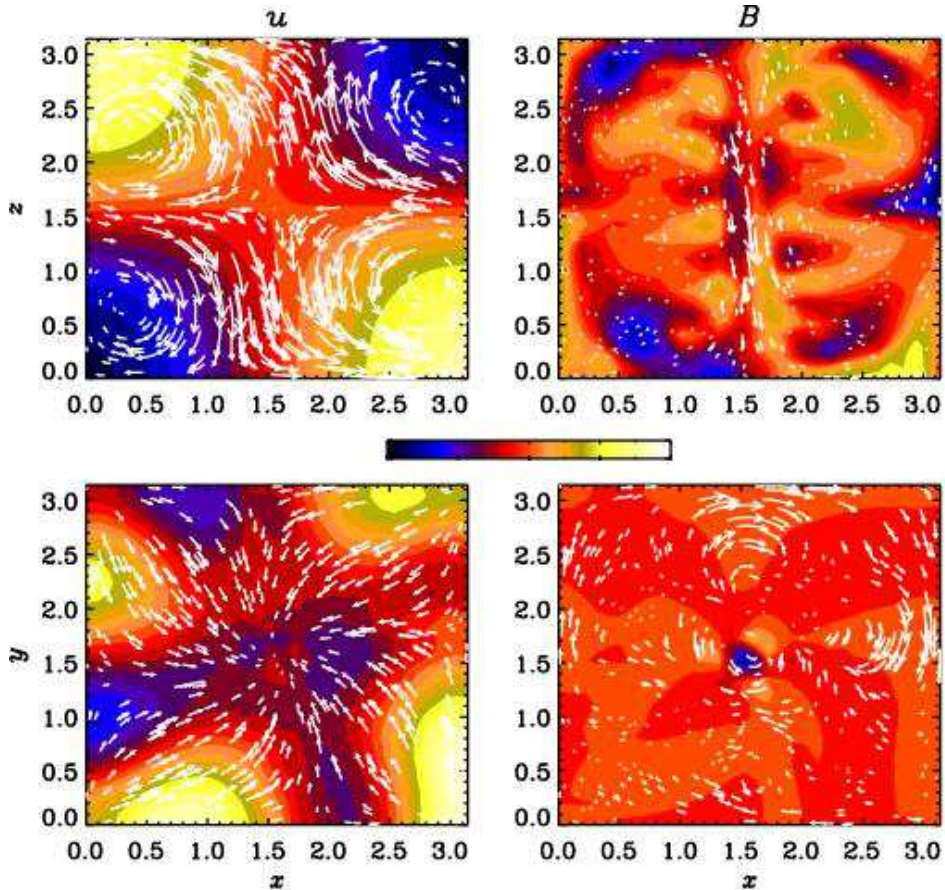}
\end{center}\caption[]{
Cross-section through the middle of the domain ($y=\pi/2$ in the upper
row and $z=\pi/2$ in the lower row) of velocity (left hand column)
and magnetic field (right hand column).
Velocity and magnetic fields are represented by vectors in the plane
and colors (gray scales) indicate the normal components.
The color (gray scale) bar indicates the contour range symmetrically
about zero within the maximum values of either sign.
The light red color (intermediate gray) corresponds to zero.
All boundaries are assumed to be {\it perfectly conducting}.
Re=5 and $R_{\rm m}=1000$.
}\label{perf32f}\end{figure}

\begin{figure}[t!]\begin{center}
\includegraphics[width=.8\columnwidth]{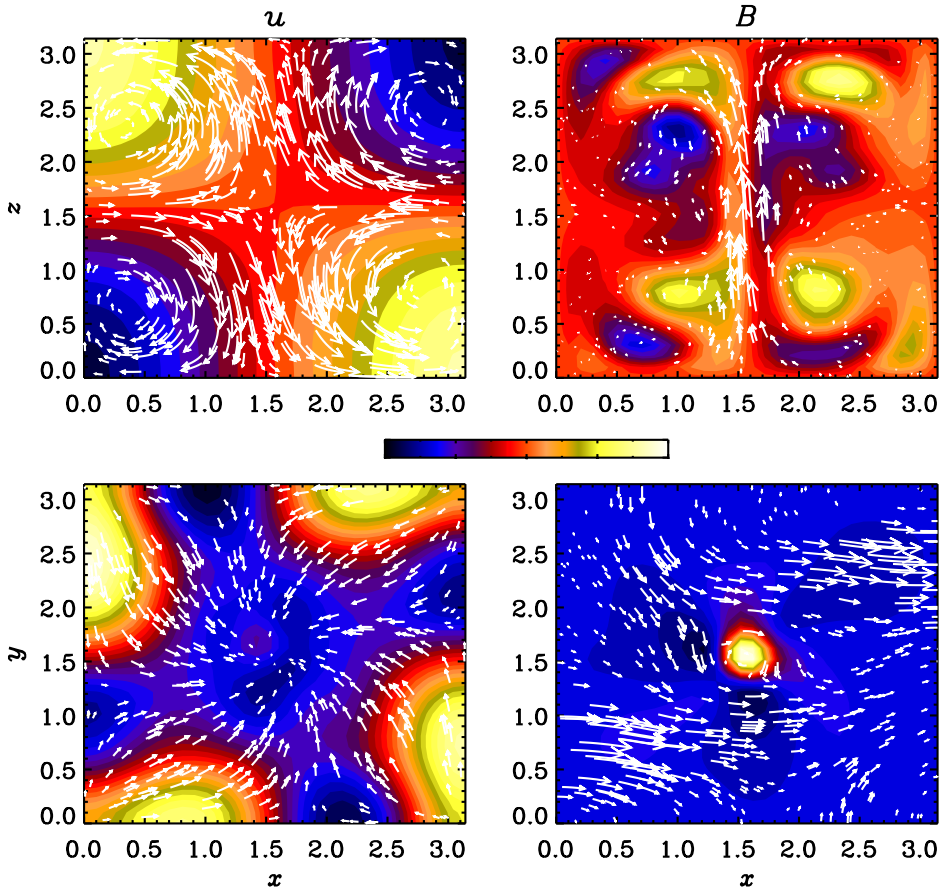}
\end{center}\caption[]{
Same as \Fig{vert}, but with {\it vertical field} boundary conditions at
$z=0$ and $z=\pi$.
Re=5 and $R_{\rm m}=200$.
}\label{vert}\end{figure}

\begin{figure}[t!]\begin{center}
\includegraphics[width=.8\columnwidth]{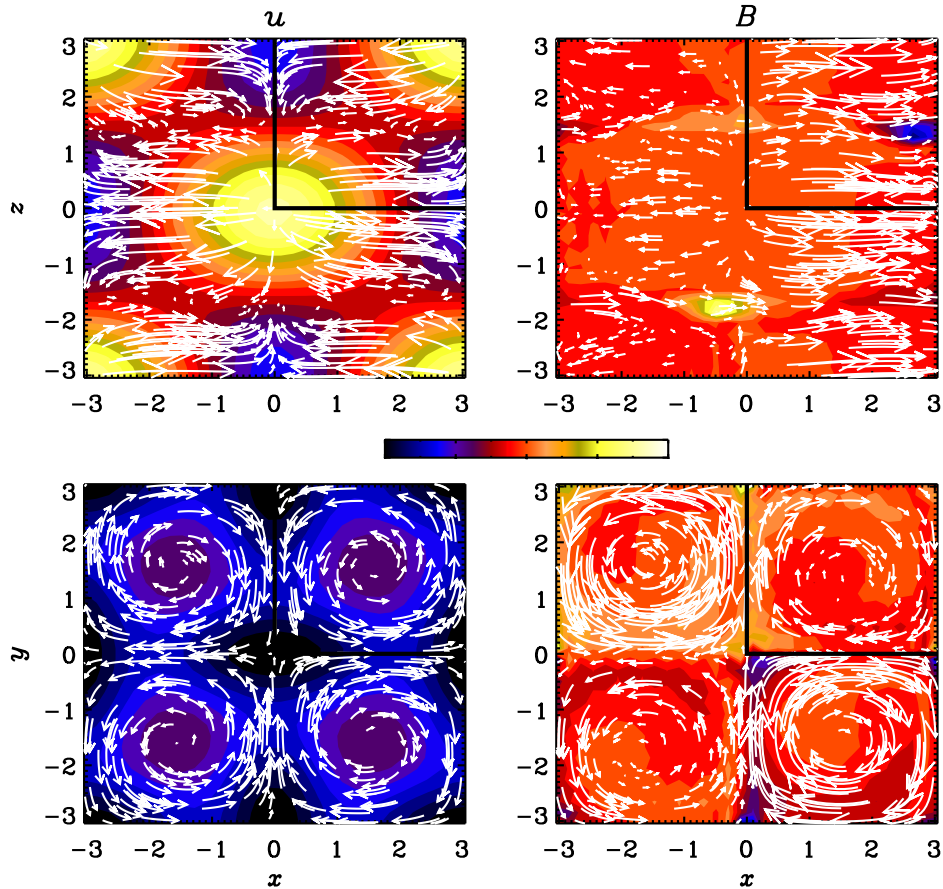}
\end{center}\caption[]{
Same as \Fig{vert}, but for triply periodic boundary conditions
in the larger domain, $-\pi<x,y,z<\pi$.
The box in the first quadrant marks the location of the computational
domains used for the results shown in \Figs{perf32f}{vert}.
Note that the magnetic field is concentrated preferentially along
the cell boundaries.
Re=5 and $R_{\rm m}=200$.
}\label{peri32c}\end{figure}

In the following we use $f_0=0.006$, $\nu=0.02$, so that $u_{\rm rms}=0.1$.
We adopt units of length where $k_0=1$.
In agreement with common practice, we use in this section Reynolds
numbers based on the scale $1/k_0$ rather than $1/k_{\rm f}$, i.e.\
$\mbox{Re}=u_{\rm rms}/(\nu k_1)$ and $R_{\rm m}=u_{\rm rms}/(\eta k_1)$.
No magnetic forcing is applied, i.e.\ $\EEE_{\rm ext} = 0$.
The VKS flow corresponds to one eight of the full domain, i.e.\
$0 < x,y,z < \pi$.
Since dynamo action in the TG flow is normally studied in triply periodic
domains we show in \Fig{perf32f} numerically obtained results for
a domain bounded by perfect conductors on all sides.
In \Fig{vert} we give the corresponding result for a domain where we adopt
a `vertical field' boundary condition on $z=0$ and $z=\pi$.
The latter is supposed to simulate the experimental situation of soft iron
boundaries with a large permeability (Fauve \& P\'etr\'elis 2003,
Morin 2005, E.\ Dormy, private communication;
see also Kenjerescaron \& Hanjalicacute 2007).
Finally, in \Fig{peri32c} we show the result for a triply periodic domain.
The resulting saturation field strengths for the three cases are
summarized in \Tab{TabTG}.

\begin{table}[htb]\caption{
Saturation values of $B_{\rm rms}=\bra{\BB^2}^{1/2}$ for TG flow dynamos
in triply-periodic domains of volume $(2\pi)^3$ (second column) compared
with those in one eighth of the volume ($\pi^3$), where the boundaries
in the $x$ and $y$ directions are perfectly conducting, and those in the
$z$ direction either using a vertical field boundary condition (third 
column) or also perfectly conducting (last column).
}\vspace{12pt}\centerline{\begin{tabular}{cccc}
$R_{\rm m}$ & periodic  &  vert.\ field & perfect cond.\ \\
\hline
 200 &    0    & 0.027 &    0   \\
 500 &  0.053  & 0.069 &    0   \\
1000 &         &       &  0.041 \\
2000 &         &       &  0.054 \\
\label{TabTG}\end{tabular}}\end{table}

Evidently, the dynamo operation is quite sensitive to
the boundary conditions in that the critical value of $R_{\rm m}$ is about
5 times larger when all domain boundaries are perfectly conducting.
In this light it is not too surprising that no dynamo was found before
using the soft iron lids at the two ends of the VKS experiment
(Monchaux et al.\ 2007).
Indeed, the model with the vertical field condition has the lowest
critical magnetic Reynolds number of all three cases considered.
The field shows a narrow vertical flux tube in the middle of the domain.
Obviously, given that our fluid Reynolds number is small, the velocity
field is laminar.
Larger Reynolds numbers could be achieved with more meshpoints.

\begin{figure}[t!]\begin{center}
\includegraphics[width=.8\columnwidth]{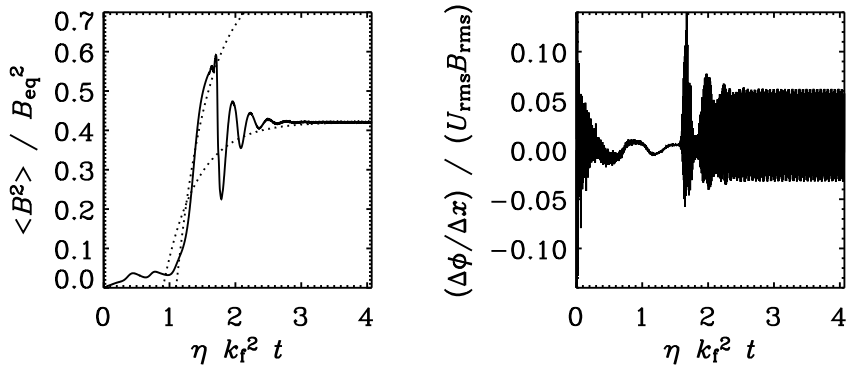}
\end{center}\caption[]{
Saturation behavior of a run with a vertical field boundary condition
similar to that shown in \Fig{vert} with $\mbox{Re}=5$ and $R_{\rm m}=500$.
The dotted lines represent failed attempts to match the saturation
behavior to the functional form of \Eq{SlowSat}, so the saturation
is not resistively limited in this case.
Note the bursty oscillations in the electric potential difference during
early times when the field becomes strong.
}\label{pubrms}\end{figure}

In \Fig{pubrms} we show that the root mean square magnetic field
strength saturates after about one magnetic diffusion time.
Nevertheless, there is {\it no} prolonged saturation phase as it was seen
in \Fig{pWeylLorenz}, or in other forced turbulence simulations where
the forcing wavenumber $k_{\rm f}/k_1$ is large.
In \Fig{pubrms} we also show the electric potential difference over
a distance of 4 meshpoints.
It turns out that the electric potential difference shows strong
bursts during times when the field is strong.
At late times the potential difference oscillates with a frequency
of about $0.14$, which is about 7 times smaller than the basic
frequency, $c_\phi k_1$.

\begin{figure}[t!]\begin{center}
\includegraphics[width=.8\columnwidth]{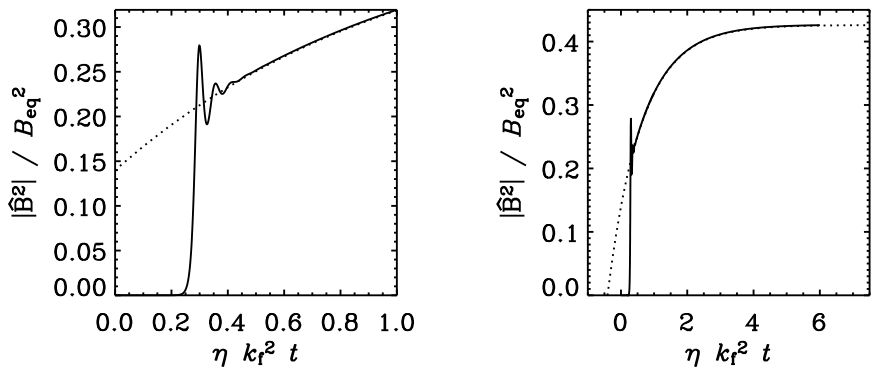}
\end{center}\caption[]{
Solution of the one-mode truncation of the Blackman--Field model
with $R_{\rm m}=500$ and $\mbox{St}=20$.
The left hand panel more clearly shows the oscillations around the
time $\eta k_{\rm f}^2 t=0.35$, while the right hand panel shows the
full saturation phase, which is matched perfectly by \Eq{SlowSat}
as shown by the dotted line.
}\label{pcomp}\end{figure}

The nature of the oscillations in \Fig{pubrms} remains unclear.
Similar oscillations have been seen in solutions of mean-field dynamo
equations derived under the $\tau$ approximation where the explicit
time derivative of the electromotive force is retained (Blackman \&
Field 2002).  Such oscillations have also been seen in direct
simulations of passive scalar turbulence, when the forcing scale is
close to the scale of the domain (Brandenburg et al.\ 2004).  This is
actually the case here, because $k_{\rm f}/k_1=\sqrt{3}$ is close to
unity.  However, there appear to be two problems with this
interpretation, which are illustrated using a numerical solution of an
appropriately adjusted version of the Blackman--Field model; see
\Fig{pcomp}.  This model is here solved in the one-mode truncation;
see \Sec{OneMode}.  Firstly, the frequency of the oscillations is in
the model $\omega_{\rm osc}=u_{\rm rms}k_{\rm f}/\sqrt{3}$, but this
is about 20 times higher than what is actually seen.  Secondly, the
slow saturation phase, as described by \Eq{SlowSat}, should still be
seen in the saturation of the mean field.  For these reasons we can
conclude that the behavior seen in Taylor-Green flow dynamos is not
described by standard mean field models.
It should also be noted that any mean
field in these dynamos cannot be seen as spatial averages.

\begin{figure}[t!]\begin{center}
\includegraphics[width=.7\columnwidth]{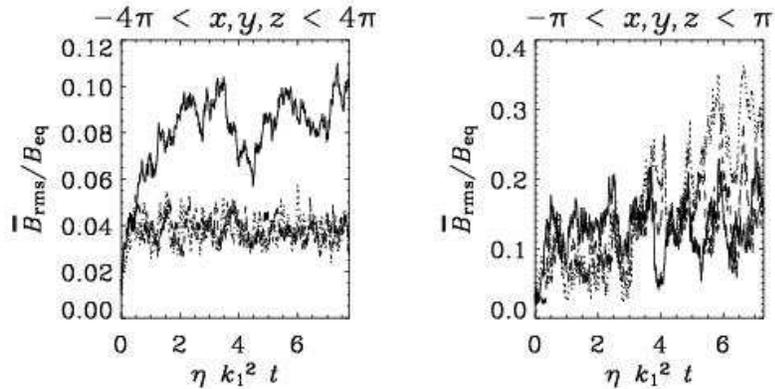}
\end{center}\caption[]{
Comparison of the evolution of the rms values of the
mean magnetic field obtained by
averaging over $xy$ planes (solid lines), $yz$ planes (dotted lines),
and $xz$ planes (dashed lines) for domains of different size.
Only in the case of a bigger domain, $-4\pi<x,y,z<4\pi$, is there
a weak mean field (solid curve).
In the case of the smaller domain, $-\pi<x,y,z<\pi$, there is in principle
also a mean field, but it it noisy and there is no preferred plane of
averaging.
In both cases $\nu=\eta=2\times10^{-4}$.
}\label{pbmzcomp}\end{figure}

Temporal averages may be a useful way of analyzing the evolution of
mean fields in Taylor-Green flow dynamos.
It turns out that after saturation ($\eta k_{\rm f}^2 t>3$) the time
averaged field contains about 88\% of the total magnetic energy.
At earlier times the oscillations are well reproduced by time averages
over short enough time spans ($\eta k_{\rm f}^2\Delta t=0.04$).
No evidence for a slow resistively limited saturation phase is seen
even for intermediate time averages.
Again, this supports the idea that the magnetic fields seen in
Taylor-Green flow dynamos cannot be described by the existing mean field
approach.

In hindsight the absence of a slow resistively limited saturation phase
is not too surprising because the wavenumber of the
forcing function, $k_{\rm f}=\sqrt{3}k_0$, is close to the box wavenumber,
so there is no scale separation.
This means that any large scale field, if it exists, could only be
generated at almost the same scale.
More importantly, because on short time scales no magnetic helicity
can be generated, the positive magnetic helicity in the forcing must
be balanced by the negative magnetic helicity at the same scale,
leading essentially to a cancellation of the magnetic field.

In order for there to be a large scale field, it is important to allow
for a domain size that is sufficiently large to accommodate a field whose
scale is at least a few times bigger than the eddy scale.
In \Fig{pbmzcomp} we show the growth of the rms field strength of
mean fields defined by averaging over different horizontal planes.
Most interesting is the $xy$ plane, and the corresponding averages are
denoted by solid lines.
Only in the case of the larger domain a mean field can be identified.
In all other cases the resulting averages are dominated by `noise'.
Nevertheless, even in the case of the bigger domain the amplitude of the
mean field is small compared with the equipartition field strength,
$B_{\rm eq}=\bra{\mu_0\rho\uu^2}^{1/2}$.

\section{Inverse transfer from a localized source}

Given that the problem of not seeing a large scale field with a
prolonged saturation phase of
$\bra{\meanBB^2}$ is related to the lack of scale separation, we now
devise a special experiment that can probably also easily be realized
in the laboratory.
The main idea is that the inverse transfer from small to large scales
is primarily connected with the conservation of magnetic helicity
(Frisch et al.\ 1975), so it is a magnetic phenomenon that is best
demonstrated using magnetic forcing in the induction equation.
This was done in the seminal paper by Pouquet et al.\ (1976) and the
resistively slow build-up of magnetic energy has also been studied by
Brandenburg et al.\ (2002; see their Fig.~12).

\begin{figure}[t!]\begin{center}
\includegraphics[width=\columnwidth]{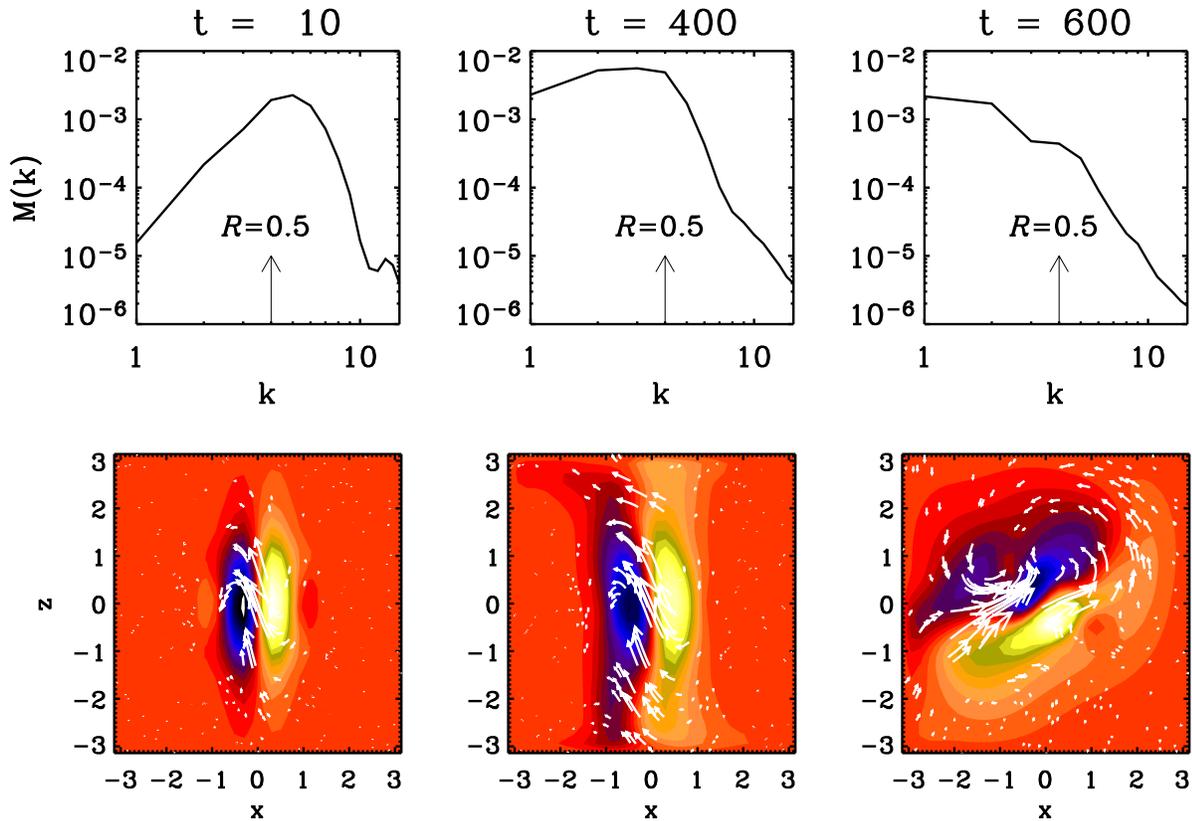}
\end{center}\caption[]{
Evolution of magnetic energy spectra and slices of the magnetic field,
driven by a localized helical electromotive force at the center with
$\sigma=1$ and $e_0=0.02$.
The vertical arrows indicate the effective forcing wavenumber,
as obtained from \Eq{kpeak}.
Here, $\nu=\eta=5\times10^{-3}$, and so the three times shown in
the figure correspond to $\eta k_1^2 t=0.05$, 2, and 3.
}\label{pspec}\end{figure}

\begin{figure}[t!]\begin{center}
\includegraphics[width=.8\columnwidth]{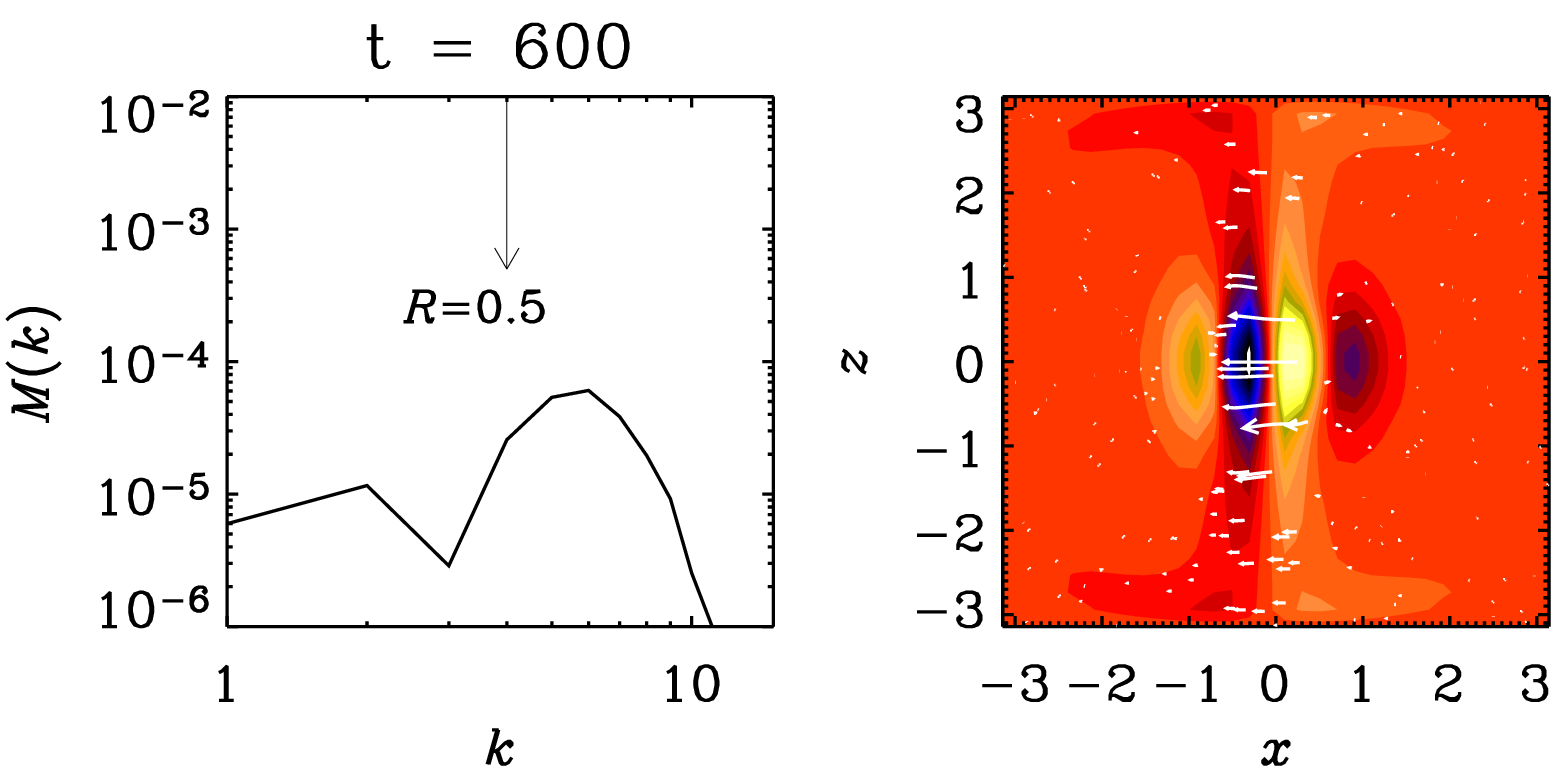}
\end{center}\caption[]{
Magnetic energy spectrum and a slice of the magnetic field
at $t=600$ for the case of a non-helical localized steady forcing
at the center with $\sigma=0$ and $e_0=0.005$.
Again, $\nu=\eta=5\times10^{-3}$, and so the time shown in
the figure corresponds to $\eta k_1^2 t=3$.
}\label{pspec_nohel}\end{figure}

The forcing needs to be helical in order to produce magnetic helicity.
In order to have scale separation we apply the forcing only in small
localized parts of the domain.
Here we restrict ourselves to only a {\it single} localized source that has
been modeled as
\EQ
\EEE_{\rm ext}(\xx)=\sigma\nab\times(\varphi\zzz)
+\nab\times\nab\times(\varphi\zzz),
\EN
where we have chosen a gaussian profile for $\varphi$ with
\EQ
\varphi(\xx)=e_0\exp\left(-\xx^2/R^2\right),
\EN
and $\sigma=1$ is chosen for a helical flow.
This yields
\EQ
\EEE_{\rm ext}(\xx)=
\pmatrix{-2\sigma y/R^2+4xz/R^4\cr 2\sigma x/R^2+4yz/R^4\cr
4(R^2-x^2-y^2)/R^4}\exp(-r^2/R^2).
\EN
Localized forcing functions of this form yield a peak in the spectrum at
\EQ
k_{\rm peak}=2/R;
\label{kpeak}
\EN
see also Mee \& Brandenburg (2006), where a potential momentum forcing
proportional to $\nab\varphi$ was adopted.
For all cases presented here we have chosen $R=0.5$, so $k_{\rm peak}=4$.
The forcing amplitude is varied between $e_0=0.005$ and 0.02 such that
the maximum flow and Alfv\'en speeds remain subsonic.

Looking at \Fig{pspec} we see that a magnetic eddy is produced that
begins to swell up until it loses its original up-down orientation and
tilts sideways in a somewhat irregular manner.
At the same time the overall magnetic energy has decreased somewhat,
but most of the spectral energy is now at large scales.
This type of behavior is not seen without helicity injection, i.e.\
for $\sigma=0$, as is demonstrated in \Fig{pspec_nohel}.

\section{Connection with the solar dynamo problem}

We now wish to discuss the possible effects of magnetic helicity
conservation that could be important for the solar dynamo.
Obviously, only pieces of this question can be addressed at this point.
Much of this has been discussed elsewhere
(e.g.\ Blackman \& Brandenburg 2002, 2003) and is summarized in various
reviews (e.g.\ Brandenburg et al.\ 2002, Brandenburg \& Subramanian 2005a).
One would generally expect magnetic helicity conservation to be
important for affecting the cycle amplitude unless magnetic helicity
is allowed to escape the dynamo domain.

Shear clearly plays an important role in dynamo processes.
Two distinct effects can be identified.
On the one hand shear can lead to so-called $\alpha\Omega$ dynamo
action which is often oscillatory.
On the other hand, shear can allow local magnetic and current helicity
fluxes along lines of constant angular velocity (Vishniac \& Cho 2001,
Subramanian \& Brandenburg 2004, 2006).
We focus here on the former effect.
In a closed domain magnetic helicity conservation acts as to produce a
``counterproductive'' alpha effect that can saturate the dynamo,
but it would not change the cycle frequency unless the turbulent
magnetic diffusivity was also catastrophically affected.
There has so far not been any clear evidence for
catastrophic $\eta_{\rm t}$ quenching.
Thus, we are at present not able to come to a conclusive results,
which is mainly a result of limited computing resources available.
However, it is also clear that in all cases the magnetic field is in
strong excess of the kinetic energy, and so any quenching would be
dominated by classical (non-catastrophic) effects.

\subsection{Catastrophic $\eta_{\rm t}$ quenching?}

For a saturated $\alpha\Omega$ dynamo the cycle frequency is equal to
$\omega_{\rm cyc}=\eta_{\rm T}k_1^2$ (Blackman \& Brandenburg 2002),
where $\eta_{\rm T}=\eta_{\rm t}+\eta$ is the total (turbulent
plus microscopic) magnetically quenched diffusivity and $k_1$
is the largest wavenumber of the domain.
This property is the key to what is perhaps the most robust method
for determining the possible dependence of $\eta_{\rm t}$ on the
magnetic field.

This was already attempted in Brandenburg et al.\ (2002)
using simulations with a sinusoidal shear profile, i.e.\
$\meanU_y=U_0\sin k_1 x$.
However, the result was not completely conclusive, because the solutions
developed some additional complexity which resulted partly from the
fact that shear strength and sign changed with $x$.
In order to avoid this problem we consider now a linear shear profile
of the form
\EQ
\meanUU=(0,Sx,0)^T,
\label{ShearFormula}
\EN
where $S$ is the shear rate.
We solve the governing equations using a shearing box approach
with shearing-periodic boundary conditions; see Hawley et al.\ (1995)
and Brandenburg et al.\ (1995).

\begin{figure}[t!]\begin{center}
\includegraphics[width=.7\columnwidth]{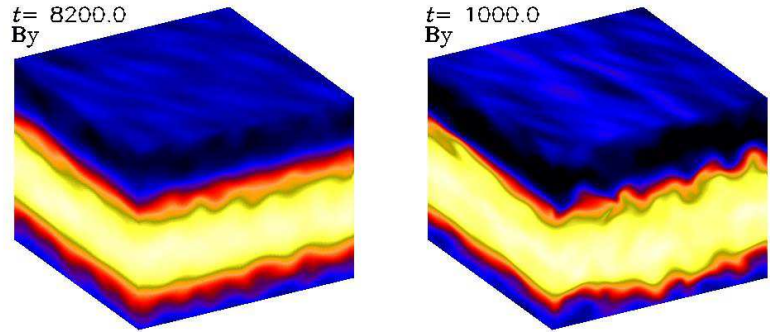}
\end{center}\caption[]{
Visualization of the toroidal field $B_y$ on the periphery of the
shearing box in the presence of forced turbulence for shear strengths
$S=-0.2$, $\nu=5\times10^{-3}$, and two different values of $\eta$,
$5\times10^{-3}$ on the left and $5\times10^{-4}$ on the right.
}\label{m64a3m256b2}\end{figure}

\begin{figure}[t!]\begin{center}
\includegraphics[width=.7\columnwidth]{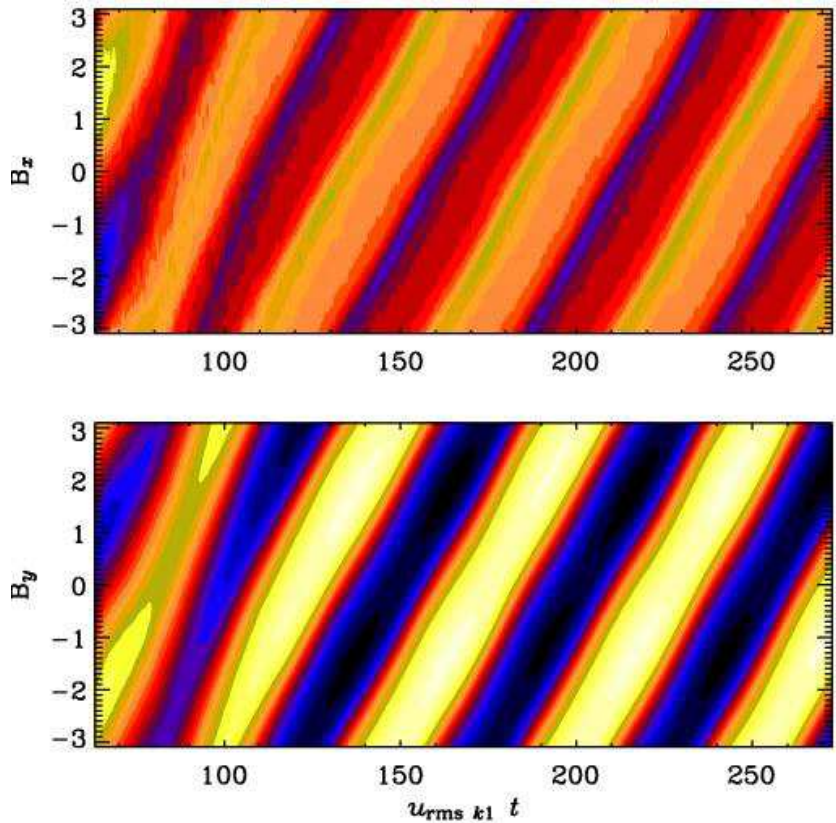}
\end{center}\caption[]{
Space-time, butterfly, or simply $zt$ diagram of $\meanB_x(z,t)$ and
$\meanB_y(z,t)$ for forced turbulence with shear, with
$S=-0.2$, $\nu=5\times10^{-3}$ and $\eta=5\times10^{-3}$.
}\label{ppaver}\end{figure}

\begin{table}[htb]\caption{
Summary of runs with uniform shear and different values of the magnetic
diffusivity and either constant kinematic viscosity or constant magnetic
Prandtl number.
}\vspace{12pt}\centerline{\begin{tabular}{cccccccc}
$\nu$ & $\eta$ & $\omega_{\rm cyc}$ & $\eta_{\rm t}$ & $\eta_{\rm t}/\eta$ &
$|\meanBB|/B_{\rm eq} $\\
\hline
$5\times10^{-3}$ & $2\times10^{-3}$ & $3.2\times10^{-3}$ & $1.2\times10^{-3}$ & 0.6 & 10 \\
$5\times10^{-3}$ & $1\times10^{-3}$ & $2.0\times10^{-3}$ & $1.0\times10^{-3}$ & 1.0 & 13 \\
$5\times10^{-3}$ & $5\times10^{-4}$ & $1.2\times10^{-3}$ & $0.7\times10^{-3}$ & 1.4 & 16 \\
$5\times10^{-3}$ & $2\times10^{-4}$ & $0.7\times10^{-3}$ & $0.5\times10^{-3}$ & 2.5 & 18 \\
\hline
$5\times10^{-3}$ & $5\times10^{-3}$ & $5.2\times10^{-3}$ & $0.2\times10^{-3}$ & 0.04& 7.6\\
$2\times10^{-3}$ & $2\times10^{-3}$ & $3.0\times10^{-3}$ & $1.0\times10^{-3}$ & 0.5 & 14 \\
\label{TabShear}\end{tabular}}\end{table}

The result of such simulations is shown in \Fig{m64a3m256b2}
for $S=-0.2$, $\nu=5\times10^{-3}$, and $\eta=2\times10^{-3}$
at a resolution of $64^3$ (on the left) and $\eta=5\times10^{-4}$
at a resolution of $256^3$ (on the right).
In both cases the field is cyclic corresponding to an upward traveling
wave, and the times are chosen such that the magnetic field is in
approximately the same phase in both figures.
The fact that the dynamo wave travels in the positive $z$ direction
is well understood as a consequence of a negative effective $\alpha$
(due to positive kinetic helicity in the forcing) and negative shear.
The traveling wave behavior is best seen in a space-time diagram
shown in \Fig{ppaver} for the simulation with $\eta=2\times10^{-3}$.
It turns out that $\omega_{\rm cyc}=0.0032$, and since $k_1=1$ this
means $\eta_{\rm T}=0.0032$ and hence $\eta_{\rm t}=0.0012$ or
$\eta_{\rm t}/\eta=0.6$.
This is still quite small, but we will be continuing to run models
at larger resolution to increase the value of $\eta_{\rm t}/\eta$.
In \Tab{TabShear} we have summarized these parameters also for a few
other runs.

\subsection{Catastrophic $\alpha$ quenching and non-locality}

The basic mechanism of catastrophic $\alpha$ quenching is now
reasonably well understood.
The basic recipe is this: whatever the mean turbulent electromotive
force $\meanEMF$ is, it leads not only to the production of large
scale magnetic fields via the standard mean field equation,
\EQ
{\partial\meanBB\over\partial t}
=\nab\times\left(\meanUU\times\meanBB+\meanEMF-\eta\mu_0\meanJJ\right),
\label{DynEqn}
\EN
but it also leads to the production of a magnetic alpha effect,
$\alpha_{\rm M}$, which characterizes the production of internal
twist in the system.
Its governing equation is
\EQ
{\partial\alpha_{\rm M}\over\partial t}+\nab\cdot\meanFF
=-2\eta_{\rm t}k_{\rm f}^2\left({\meanEMF\cdot\meanBB\over B_{\rm eq}^2}
+{\alpha_{\rm M}\over R_{\rm m}}\right).
\label{QuenEqn}
\EN
Here we have allowed for the possibility of an additional flux of
small scale magnetic helicity.
Also, in this equation $R_{\rm m}$ is meant to represent
$\eta_{\rm t}/\eta$ by definition; see Blackman \& Brandenburg (2002)
for details.
Note that the magnetic helicity equation is unaffected by the large scale
velocity term, $\meanUU\times\meanBB$.

It is important to realize that the possibility of catastrophic
quenching is quite general and not restricted to the local
$\alpha$ effect formula considered here.
In some recent solar dynamo models the so-called Babcock-Leighton
mechanism is used (e.g.\ Dikpati \& Charbonneau 1999).
Here one assumes that there is some source term to the electromotive
force that is localized near the surface, but it is proportional
to the toroidal field a the bottom of the convection zone.
This effect is sometimes thought of being distinct from the usual $\alpha$
effect in that it allows for the generation of super-equipartition
field strengths.

Formally, the Babcock-Leighton mechanism is just a non-local $\alpha$
effect where the multiplication in $\alpha\meanBB$ is replaced by a
convolution, $\hat\alpha\circ\meanBB$, where $\hat\alpha$ is an
integral kernel.
Applying the same idea also to the turbulent magnetic diffusivity
$\hat\eta_{\rm t}(z,z')$, leads to an expression for $\meanEMF(z,t)$
of the form
\EQ
\meanEMF(z,t)=\int_{z_1}^{z_2}\hat\alpha(z,z')\meanBB(z',t)\,\dd z'
-\int_{z_1}^{z_2}\hat\eta(z,z')\meanJJ(z',t)\,\dd z'.
\EN
The possibility of nonlocal $\alpha$ and $\eta$ effects has been inferred
also from simulation data of magneto-rotational turbulence in accretion
discs (Brandenburg \& Sokoloff 2002).
Nonlocal $\alpha$ effects are conveniently characterized in terms of
its spectral decomposition,
\EQ
\tilde\alpha(k)=\int_{z_1}^{z_2}\sin kz\,\sin kz'\;
\hat\alpha(z,z')\,\dd z\,\dd z'.
\label{SpectralAlpha}
\EN
This technique is particularly convenient because in Fourier space the
convolution corresponds to a multiplication.

\begin{figure}[t!]\begin{center}
\includegraphics[width=.7\columnwidth]{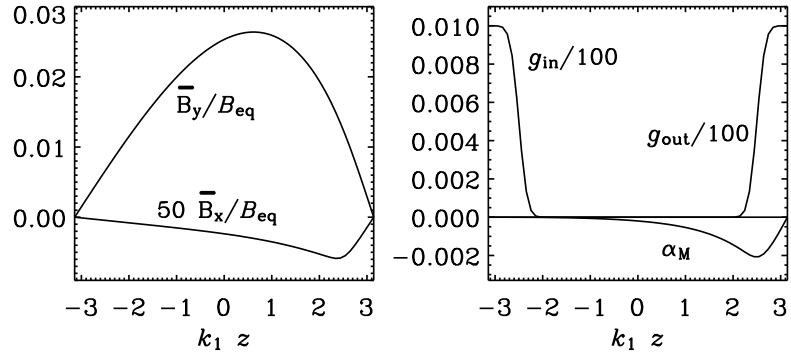}
\end{center}\caption[]{
Profiles of the magnetic field and the magnetic contribution to the $\alpha$
effect for the nonlocal dynamo model with $R_{\rm m}=10^3$.
}\label{pprofs}\end{figure}

\begin{figure}[t!]\begin{center}
\includegraphics[width=.5\columnwidth]{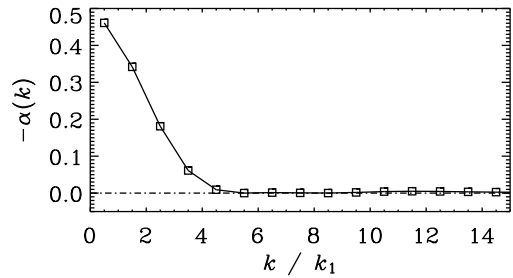}
\end{center}\caption[]{
Profile of the $\alpha$ kernel for the nonlocal (Babcock-Leighton type)
dynamo model computed using \Eq{SpectralAlpha}.
}\label{palpk}\end{figure}

In the following we restrict ourselves to a simple expression of the form
\EQ
\hat{\alpha}(z,z')=\alpha_0\, g_{\rm out}(z)\, g_{\rm in}(z'),
\EN
where
\EQ
g_{\rm out}(z)=\half\left[1+\erf\left({z-z_2\over d}\right)\right],
\EN
\EQ
g_{\rm in}(z')=\half\left[1-\erf\left({z'-z_1\over d}\right)\right]
\EN
are simple profile functions representing the peak of the source function
near $z=z_2$ with a sensitivity for fields located near $z=z_1$.
For the following we choose $-z_1=z_2=2.5$ and $d=0.05$ in the domain
$-\pi<z<\pi$; see the right hand panel of \Fig{pprofs}.
The resulting profile of $\tilde\alpha(k)$ according to \Eq{SpectralAlpha}
is shown in \Fig{palpk}.

\begin{figure}[t!]\begin{center}
\includegraphics[width=.7\columnwidth]{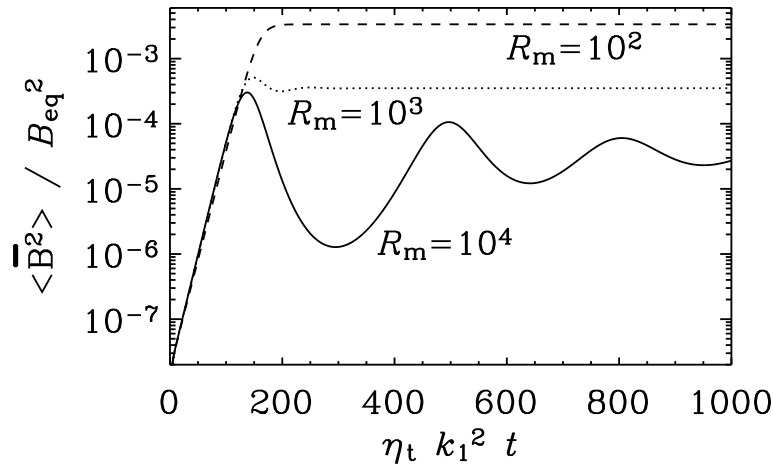}
\end{center}\caption[]{
Evolution of the magnetic energy of the large scale field for
$R_{\rm m}$ between $10^2$ and $10^4$ for the nonlocal dynamo model.
For large of $R_{\rm m}$ the linear growth rate reaches an
asymptotic value, but the nonlinear saturation amplitude continues
to depend on $R_{\rm m}$.
}\label{pcomp2}\end{figure}

We solve \Eqs{DynEqn}{QuenEqn} numerically for the case $\meanFF=0$
using an implicit scheme for $\alpha_{\rm M}$ by writing the equation
in the form
\EQ
{\alpha_{\rm M}^{n+1}\!-\alpha_{\rm M}^{n}\over\delta t}
+2\eta_{\rm t}k_{\rm f}^2{\meanEMF_0\cdot\meanBB\over B_{\rm eq}^2}
+\eta k_{\rm f}^2\left(1+R_{\rm m}{\meanBB^2\over B_{\rm eq}^2}\right)
\left(\alpha_{\rm M}^{n+1}+\alpha_{\rm M}^{n}\right)=0,\;
\label{QuenEqnNum}
\EN
where $\meanEMF_0=\meanEMF-\alpha_{\rm M}\meanBB$ is the electromotive
force without the magnetic quenching term.
\EEq{QuenEqnNum} is then solved for $\alpha_{\rm M}^{n+1}$ at the new time,
$t_{n+1}=t_n+\delta t$, using for $\meanBB$ the value at the present time
level $t_n$, i.e.\
\EQ
\alpha_{\rm M}^{n+1}=(1+Q)^{-1}\left[(1-Q)\,\alpha_{\rm M}^{n}
-2\eta_{\rm t}k_{\rm f}^2\delta t{\meanEMF_0\cdot\meanBB
\over B_{\rm eq}^2}\right],
\label{QuenEqnNum2}
\EN
where $Q=\eta k_{\rm f}^2\delta t(1+R_{\rm m}\meanBB^2/B_{\rm eq}^2)$.
We have considered a model using linear shear of the form \eq{ShearFormula}.
The strength of shear and $\alpha$ effect are quantified by the
non-dimensional numbers
\EQ
C_S={S\over\eta_{\rm t}k_1^2},\quad
C_\alpha={\alpha\over\eta_{\rm t}k_1},
\EN
where $k_1=1$ is the smallest wavenumber in the computational domain.
In the following we use $C_S=100$, $C_\alpha=0.1$, and $k_{\rm f}/k_1=5$.
In many cases an explicit treatment of the $\alpha_{\rm M}$ equation
suffices (e.g.\ Blackman \& Brandenburg 2002), but in the present case
an explicit solution algorithm was found to be unstable.

It turns out that most of the field is generated in the middle of the
domain, while most of the quenching via $\alpha_{\rm M}$ occurs near
the top layers around $z=z_2$; see \Fig{pprofs}.
Nevertheless, as expected, this model experiences still catastrophic
quenching; see \Fig{pcomp2}.
These results look quite similar to those obtained for local $\alpha$
profiles (Brandenburg \& Subramanian 2005b).
We note that for $R_{\rm m}=10^4$ it is important to perform the
calculations using double precision arithmetics.

\subsection{Location of the dynamo}

The location of the solar dynamo is rather uncertain.
Particularly unclear is the location where most of the toroidal
field resides.
The standard thinking since the 1980s is that most of the
toroidal field can only reside at the base of the convection zone,
because magnetic buoyancy would remove the field on a short time scale.
However, it turned out that downward pumping of magnetic field helps to 
keep the magnetic field inside the convection zone.
On the other hand, the toroidal field has been argued to be actually
quite strong, such that it would exceed the equipartition field
strength by factors as large as 100. 
Furthermore, stability of such strong fields even beneath the
convection zone have recently been put under doubt (Arlt et al.\ 2007,
Kitchatinov \& R\"udiger\ 2007).
Whether or not such strong fields could be generated by a turbulent
dynamo is unclear.
This led to the proposal that the solar dynamo may instead be located
in the bulk of the convection zone and that the mean fields are at most
comparable in strength to the equipartition field strength
(Brandenburg 2005).

Astrophysical dynamos are not expected to be catastrophically quenched,
because they are likely to shed excess small scale magnetic helicity
through magnetic helicity fluxes.
This would mean that the quenching through $\alpha_{\rm M}$ is much weaker
than the other nonlinearities, e.g.\ suppression of the mean flow or
of $\alpha_{\rm K}$ itself.
The latter can be {\it approximated} by the more conventional expression
$\alpha_{\rm K}=\alpha_{\rm K0}/(1+\BB^2/B_{\rm eq}^2)$ (for detailed
expressions see Rogachevskii \& Kleeorin 2000).

If the solar dynamo does indeed work in a distributed fashion, then the
meridional circulation is probably no longer important for determining
the cycle period and the equatorward migration of the magnetic flux belts.
Instead, the dynamo may be essentially of a standard $\alpha\Omega$ type.
However, an important problem arises from the fact that the near-surface
shear layer, where $\partial\Omega/\partial r<0$, is rather thin.
Given that the aspect ratio of magnetic flux belts is usually of order
unity, it is difficult to envisage how one can explain the rather
broad latitudinal distribution of flux of the same sign.

\begin{figure}[t!]\begin{center}
\includegraphics[width=.6\columnwidth]{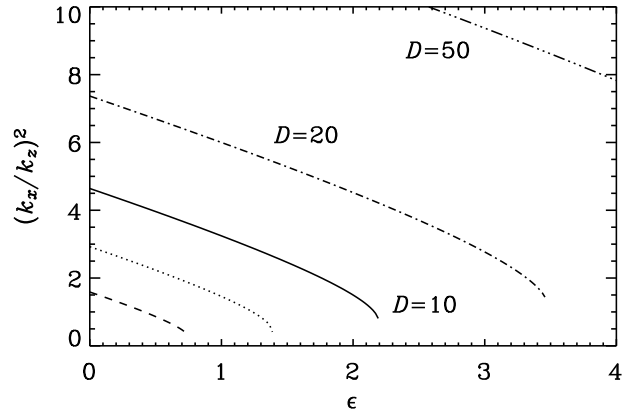}
\end{center}\caption[]{
Aspect ratio as a function of the degree of anisotropy, $\epsilon$
for $D=2$ (dashed line), 5 (dotted line), 10, 20, and 50.
When the lines stop for small values of $\kappa=k_x/k_z$
there are no solutions any more.
}\label{pkxcrit}\end{figure}

As a possible solution to this problem one might think of the effects
of magnetic helicity transport and possibly the anisotropy of the turbulent
magnetic diffusivity.
However, magnetic helicity fluxes modify primarily the effective $\alpha$
and it is not clear that they affect the aspect ratio of the toroidal
flux belts.
In the following we address the effect of an anisotropic magnetic diffusivity.
Based on the dispersion relation for a dynamo wave in a two-dimensional domain
(see, e.g., Brandenburg \& Subramanian 2005a, Sect.~6.5.2),
\EQ
\lambda=-\eta_{\rm T}(k_x^2+\epsilon k_z^2)+\sqrt{\half\alpha S k_x},
\EN
where $\epsilon$ is the degree of anisotropy, we find that in the 
marginally excited case, $\lambda=0$,
the value of $k_x$ can be obtained iteratively.
For this purpose, let us define the aspect ratio as $\kappa=k_x/k_z$,
which is then given by
\EQ
\kappa^2+\epsilon=\sqrt{\half\alpha S\kappa/\eta_{\rm T}^2k_z^3}.
\label{kappaeqn}
\EN
We define the dynamo number as
\EQ
D=\half\alpha S/\eta_{\rm T}^2k_z^3,
\EN
so \Eq{kappaeqn} reduces to
\EQ
\kappa^2=\sqrt{D\kappa}-\epsilon,
\EN
which can be solved iteratively.
The result is shown in \Fig{pkxcrit}.
In the special case $\epsilon=0$ we have $\kappa=D^{1/3}$.

It turns out that anisotropic diffusion tends to increase the ratio
$\kappa=k_x/k_z$, so the wavelength in the latitudinal (or $x$) direction
becomes smaller, making the problem even worse.
Obviously one must be careful with linear theory, so this result may
not be too meaningful, but in the absence of any other evidence there
is currently no particular reason to expect anisotropic diffusion being
a solution to the problem of the aspect ratio.

\subsection{Predictability in distributed dynamos}

Flux transport dynamos have been used for predicting the strength of the
next solar cycle, Cycle~24 (see Clarke 2006 for a general assessment).
The outcome depends essentially on the way the observed solar activity is
used to keep the evolution of the model in sync with the Sun.
In Dikpati \& Gilman (2005, 2006) the sunspot number is used as a proxy of
the poloidal field production by a source term near the surface.
On the other hand, in Choudhuri et al.\ (2007) the observed poloidal field
around the poles is used to correct the poloidal field in the upper part of
their model during solar minimum.
Regardless of the fact that the outcomes can be very different
(strong Cycle~$24$ in the former model and weak Cycle~$24$ in the latter),
it is clear that this topic has attracted significant attention
(Clarke 2006, Tobias et al.\ 2006, Cameron \& Sch\"ussler 2006) and has
led to the impression that the solar dynamo problem might be solved.

\begin{figure}[t!]\begin{center}
\includegraphics[width=.8\columnwidth]{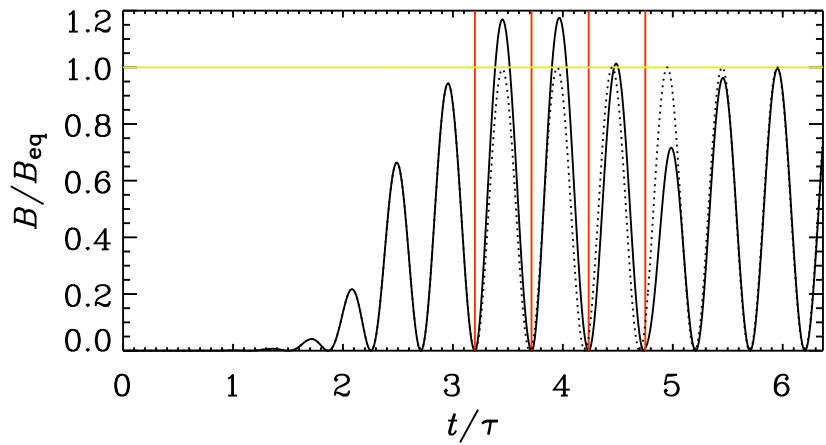}
\end{center}\caption[]{
Simple dynamo model with dynamo number $D=4$ with initial condition
$A=0.01$ and $B=0$.
The model reaches saturation at $t/\tau\approx3$.
The poloidal field is rescaled by factors 1.25, 1.23, 1.00, and 0.60
during the minima marked by vertical lines.
The unscaled model is overplotted as a dotted line.
Note the weak cycle amplitude at $t/\tau\approx5$,
corresponding to Cycle~24.
}\label{pmodel}\end{figure}

The point of this section is to show that the predictive power of a model
is not very sensitive to the details, and that even in the
extreme case of a fully distributed toy model similar predictability can
be achieved.
To demonstrate this,
we consider here the following simple model equations that are obtained
from a single mode truncation of a one-dimensional periodic model, i.e.\
\EQ
{\dd A\over\dd t}=\alpha B-\tau^{-1} A,
\EN
\EQ
{\dd B\over\dd t}=ik\Omega' A-\tau^{-1} B,
\EN
for the complex variables $A$ and $B$, that characterize the poloidal
and toroidal fields, respectively (Durney \& Robinson 1982).
Here, $\tau=(\eta_{\rm t}k)^{-1}$ is the turbulent magnetic diffusion time,
$\Omega'$ is the radial gradient of the angular velocity, and
$\alpha$ is the $\alpha$ effect.
We assume here a simple form of $\alpha$ quenching with
$\alpha=\alpha_0/(1+|B|^2/B_{\rm eq}^2)$, where $\alpha_0$ is the
kinematic value of the $\alpha$ effect, and $B_{\rm eq}$ is the
equipartition field strength.
The non-dimensional dynamo number, $D=\alpha_0\Omega' kL\tau^{-2}$
has to exceed the value $D=2$ for dynamo action.

Following Choudhuri et al.\ (2007) we use normalized values of the
observed dipole moment (as given by Svalgaard et al.\ 2005) to correct
the poloidal field amplitude $A$ by factors 1.25, 1.23, 1.00, and 0.60
after Cycles~20, 21, 22, and 23, respectively.
The result is shown in \Fig{pmodel}.
The maxima in $|B|$ after each of the four rescalings are
1.17, 1.17, 1.02, and 0.71 times the usual values.
Applied to the Sun, this means that the next cycle will indeed
be about 30\% weaker than the previous one.
This is obviously in perfect agreement with the model of
Choudhuri et al.\ (2007) and of course the earlier investigations
by Svalgaard et al.\ (2005).

\section{Conclusions}

There has been tremendous progress in dynamo theory
since the beginning of the millennium,
both theoretically and experimentally.
In this paper we do not intend to review any of this progress.
Instead, we have focussed on a number of new ideas that emerged
in an attempt to connect theory and experiments.
The relevance of experiments is not immediately obvious because in
experiments the
value of $R_{\rm m}$ is currently below 50 or so, while simulations
can reach values on the order of 1000.
On the other hand, in experiments the magnetic Prandtl number is small
($10^{-5}$), which is quite similar to the value in the Sun and other stars.
To reach interesting values of $R_{\rm m}$ that allow for dynamo action or
other effects to occur, the fluid Reynolds number has to be very large
(in excess of $10^6$ or so), which is out of reach of simulations even
in the intermediate future.

On the other hand, experiments are now beginning to produce dynamo
action in unconstrained flows (Monchaux et al.\ 2007), and corresponding
experiments are now beginning to produce results that can meaningfully
be compared with experiments.
Examples include the work of Nore et al.\ (2006) for simulating flows
close to those in the VKS experiment, and approximations to the VKS flow
in terms of Taylor-Green flows (Ponty et al.\ 2004, Mininni et al.\ 2005b).
Although neither of these cases is anywhere near the conditions relevant
to the Sun, it is conceivable that the theoretical investigations that have
been undertaken in just the last few years (e.g.\ regarding magnetic
helicity conservation and helicity fluxes) can lead to new paradigms that
can in principle be tested experimentally (e.g.\ the phenomenon of
inverse cascade-type behavior).

Even though progress is rapid, and success appears sometimes in reach,
there are still a number of problems that are poorly understood.
An example concerns the magnetic helicity flux whose effects are
manifested in coronal mass ejections (D\'emoulin et al.\ 2002),
but they have hardly been incorporated in dynamo models with the
aim to alleviate the otherwise catastrophic quenching.

\section*{Acknowledgements}

We thank Eric Blackman for making useful suggestions to an earlier
version of the manuscript.
We acknowledge the allocation of computing resources provided by the
Centers for Scientific Computing in Denmark (DCSC), Finland (CSC),
and Sweden (PDC).

\appendix
\section{Magnetic versus kinetic helicity conservation}
\label{analogy}

In this appendix we want to show why there is such a dramatic
difference between the conservation of magnetic and kinetic helicities.
The kinetic helicity is indeed conserved if there is no magnetic field,
i.e.\ no Lorentz force, and if $\nu=0$ exactly.
Let us also assume $\nab\cdot\AAA=\nab\cdot\UU=0$ for simplicity,
although this is not critical.
The pair of analogous equations can then be written as
\EQ
{\partial\AAA\over\partial t}=\UU\times\BB-\eta\JJ-\nab\phi+\EEE_{\rm ext},
\EN
\EQ
{\partial\UU\over\partial t}=\UU\times\WW-\nu\QQ-\nab p+\FFF_{\rm ext}.
\EN
For later reference we also quote here the curled evolution equations,
where $\BB=\nab\times\AAA$ is the magnetic field and
$\WW=\nab\times\UU$ is the vorticity, i.e.\
\EQ
{\partial\BB\over\partial t}
=\nab\times(\UU\times\BB)-\eta\nab\times\JJ+\nab\times\EEE_{\rm ext},
\EN
\EQ
{\partial\WW\over\partial t}
=\nab\times(\UU\times\WW)-\nu\nab\times\QQ+\nab\times\FFF_{\rm ext}.
\EN
Note that $\nab\cdot\BB=\nab\cdot\WW=0$ is automatically fulfilled.
Denoting, as usual, volume averages by angular brackets, and assuming
periodic boundaries, the evolution equations for the two helicities are
\EQ
{\partial\over\partial t}\bra{\AAA\cdot\BB}=
-2\eta\bra{\JJ\cdot\BB}+2\bra{\BB\cdot\EEE_{\rm ext}},
\EN
\EQ
{\partial\over\partial t}\bra{\UU\cdot\WW}=
-2\nu\bra{\QQ\cdot\WW}+2\bra{\WW\cdot\FFF_{\rm ext}}.
\EN
For completeness we give here also the evolution equations for the
energy norms,
\EQ
{\partial\over\partial t}\half\bra{\AAA^2}=
-\bra{\UU\cdot(\AAA\times\BB)}
-\eta\bra{\BB^2}+\bra{\AAA\cdot\EEE_{\rm ext}},
\EN
\EQ
{\partial\over\partial t}\half\bra{\UU^2}=
-\mbox{\sout{$\bra{\UU\cdot(\UU\times\WW)}$}}
-\nu\bra{\WW^2}+\bra{\UU\cdot\FFF_{\rm ext}},
\EN
\EQ
{\partial\over\partial t}\half\bra{\BB^2}=
-\bra{\UU\cdot(\JJ\times\BB)}
-\eta\bra{\JJ^2}+\bra{\JJ\cdot\EEE_{\rm ext}},
\EN
\EQ
{\partial\over\partial t}\half\bra{\WW^2}=
-\bra{\UU\cdot(\QQ\times\WW)}
-\nu\bra{\QQ^2}+\bra{\QQ\cdot\FFF_{\rm ext}}.
\EN
Note that we kept the term $\bra{\UU\cdot(\UU\times\WW)}=0$
in order to enhance the formal analogy of the equations.
In the helicity equations one could restore two similarly redundant terms,
\EQ
\half{\partial\over\partial t}\bra{\AAA\cdot\BB}=
-\mbox{\sout{$\bra{\UU\cdot(\BB\times\BB)}$}}
-\eta\bra{\JJ\cdot\BB}+\bra{\BB\cdot\EEE_{\rm ext}}
\EN
\EQ
\half{\partial\over\partial t}\bra{\UU\cdot\WW}=
-\mbox{\sout{$\bra{\UU\cdot(\WW\times\WW)}$}}
-\nu\bra{\QQ\cdot\WW}+\bra{\WW\cdot\FFF_{\rm ext}}
\EN
In conclusion, all the evolution equations have the following three terms:
an internal driving term such as $\bra{\UU\cdot(\JJ\times\BB)}$,
an external driving term such as $\bra{\UU\cdot\FFF_{\rm ext}}$,
and a viscous or resistive loss term.
In the magnetic energy equation there is the internal driving term
(work done against the Lorentz force), but usually no external driving
term, since $\EEE_{\rm ext}=0$ is assumed here.
The kinetic energy equation, on the other hand, does not have an
internal driving term, because $\bra{\UU\cdot(\UU\times\WW)}=0$.
Thus, both kinetic and magnetic energy equations have driving terms,
so we have
\EQ
{\partial\over\partial t}\half\bra{\BB^2}=
-\bra{\UU\cdot(\JJ\times\BB)}-\eta\bra{\JJ^2},
\EN
\EQ
{\partial\over\partial t}\half\bra{\UU^2}=
\bra{\UU\cdot\FFF_{\rm ext}}-\nu\bra{\WW^2}.
\EN
Thus, in a statistically steady state, the dissipative
terms have to balance the corresponding driving terms.
In the limit of large magnetic and fluid Reynolds numbers this leads
to the asymptotic scalings
\EQ
|\JJ|\sim\eta^{-1/2},\quad
|\WW|\sim\nu^{-1/2}.
\EN
At the same time, because kinetic and magnetic energies are bounded,
magnetic field strength and velocity do not diverge but stay independent
of $\eta$ and $\nu$, respectively.
This also implies that the typical inverse length scales scale like
\EQ
k\sim|\WW|/|\UU|\sim\nu^{-1/2},
\EN
and hence that
\EQ
|\QQ|\sim k|\WW|\sim\nu^{-1},
\EN
so that the magnetic and kinetic helicity dissipation terms scale like
\EQ
|\eta\bra{\JJ\cdot\BB}|\to\eta^{1/2}\to0,
\EN
\EQ
|\nu\bra{\QQ\cdot\WW}|\to\nu^{-1/2}\to\infty,
\EN
which highlights the fundamental difference between kinetic and
magnetic helicity conservation when $\eta$ and $\nu$ are not zero
but small.

\section{The forcing function}
\label{ForcingFunction}

For completeness we specify here the forcing function used in the
present paper\footnote{This forcing function was also used by
Brandenburg (2001), but in his Eq.~(5) the factor 2 in the denominator
should have been replaced by $\sqrt{2}$ for a proper normalization.}.
It is defined as
\EQ
\ff(\xx,t)={\rm Re}\{N\ff_{\kk(t)}\exp[\ii\kk(t)\cdot\xx+\ii\phi(t)]\},
\EN
where $\xx$ is the position vector.
The wavevector $\kk(t)$ and the random phase
$-\pi<\phi(t)\le\pi$ change at every time step, so $\ff(\xx,t)$ is
$\delta$-correlated in time.
For the time-integrated forcing function to be independent
of the length of the time step $\delta t$, the normalization factor $N$
has to be proportional to $\delta t^{-1/2}$.
On dimensional grounds it is chosen to be
$N=f_0 c_{\rm s}(|\kk|c_{\rm s}/\delta t)^{1/2}$, where $f_0$ is a
nondimensional forcing amplitude.
At each timestep we select randomly one of many possible wavevectors
in a certain range around a given forcing wavenumber.
The average wavenumber is referred to as $k_{\rm f}$.
We force the system with transverse helical waves,
\begin{equation}
\ff_{\kk}=\RRRR\cdot\ff_{\kk}^{\rm(nohel)}\quad\mbox{with}\quad
{\sf R}_{ij}={\delta_{ij}-\ii\sigma\epsilon_{ijk}\hat{k}_k
\over\sqrt{1+\sigma^2}},
\end{equation}
where $\sigma=1$ for positive helicity of the forcing function,
\EQ
\ff_{\kk}^{\rm(nohel)}=
\left(\kk\times\eee\right)/\sqrt{\kk^2-(\kk\cdot\eee)^2},
\label{nohel_forcing}
\EN
is a non-helical forcing function, and $\eee$ is an arbitrary unit vector
not aligned with $\kk$; note that $|\ff_{\kk}|^2=1$.

\section{One-mode truncation of the Blackman--Field model}
\label{OneMode}

A one-mode truncation of the dynamically quenched $\alpha$ effect
has been studied by Blackman \& Brandenburg (2002), where
$\meanBB(z,t)=\hatBB(t)\exp(\ii k_1 z)$ has been assumed.
Here we include the explicit time dependence of the electromotive force
$\meanEMF(z,t)=\hatEMF(t)\exp(\ii k_1 z)$, so our model equations are
\EQ
{\dd\hatBB\over\dd t}=\ii\kk_1\times(\hatEMF-\eta\hatJJ),
\EN
\EQ
{\dd\hatEMF\over\dd t}={\tilde\alpha}\hatBB-{\tilde\eta}_{\rm t}\hatJJ
-{\hatEMF\over\tau},
\EN
\EQ
{\dd\alpha_{\rm M}\over\dd t}=-2\eta_{\rm t}k_f^2\left[
{\mbox{Re}(\hatEMF^*\cdot\hatBB)\over B_{\rm eq}^2}
+{\alpha_{\rm M}\over R_{\rm m}}\right],
\EN
where $\hatBB$ and $\hatEMF$ are complex and $\alpha_{\rm M}$ a
real dependent variable, $\kk_1=(0,0,k_1)^T$ is the wave vector,
$\alpha=\alpha_{\rm K}+\alpha_{\rm M}$ with
$\alpha=\tau\tilde\alpha$ and $\eta_{\rm t}=\tau\tilde\eta_{\rm t}$,
and $\tau$ is the relaxation time.
The closure assumption consists in representing the triple correlations
by the damping term $\hatEMF/\tau$.

In order to associate this model with a simulation, we use the values of
$k_{\rm f}/k_1$, $R_{\rm m}$ and chose a value of $\mbox{St}$ to determine
the following set of model parameters:
$u_{\rm rms}=3\eta_{\rm t}k_{\rm f}/\mbox{St}$,
$\tau=\mbox{St}/(u_{\rm rms}k_{\rm f})$, and
$\alpha_{\rm K}=\eta_{\rm t} k_{\rm f}$, which assumes perfectly helical
turbulence, which is obviously not realistic for Taylor-Green flow dynamos.

\end{document}